\documentclass[twocolumn,english,aps,prl,showpacs,amsmath,amssymb,superscriptaddress]{revtex4}
\usepackage[latin9]{inputenc}
\usepackage{color}
\usepackage{amstext}
\usepackage{amssymb}
\usepackage{graphicx}

\makeatletter
\@ifundefined{textcolor}{}
{%
 \definecolor{BLACK}{gray}{0}
 \definecolor{WHITE}{gray}{1}
 \definecolor{RED}{rgb}{1,0,0}
 \definecolor{GREEN}{rgb}{0,1,0}
 \definecolor{BLUE}{rgb}{0,0,1}
 \definecolor{CYAN}{cmyk}{1,0,0,0}
 \definecolor{MAGENTA}{cmyk}{0,1,0,0}
 \definecolor{YELLOW}{cmyk}{0,0,1,0}
}

\usepackage{color}

\newcommand{\beq}{\begin{eqnarray}}
\newcommand{\eeq}{\end{eqnarray}}

\makeatother

\usepackage{babel}
\begin{document}

\title{Spin-isotropic continuum of spin excitations in antiferromagnetically ordered Fe$_{1.07}$Te}

\author{Yu Song}

\email{Yu.Song@rice.edu}

\selectlanguage{english}%

\affiliation{Department of Physics and Astronomy, Rice University, Houston, Texas
77005, USA}

\author{Xingye Lu}

\affiliation{Center for Advanced Quantum Studies and Department of Physics, Beijing
	Normal University, Beijing 100875, China}

\author{L.-P. Regnault}

\affiliation{Institut Laue Langevin, 71 Avenue des Martyrs, 38042 Grenoble, France}

\author{Yixi Su}

\affiliation{J{ü}lich Centre for Neutron Science, Forschungszentrum J{ü}lich
GmbH, Outstation at MLZ, D-85747 Garching, Germany}

\author{Hsin-Hua Lai}

\affiliation{Department of Physics and Astronomy, Rice University, Houston, Texas 77005, USA}

\author{Wen-Jun Hu}

\affiliation{Department of Physics and Astronomy, Rice University, Houston, Texas 77005, USA}	

\author{Qimiao Si}

\affiliation{Department of Physics and Astronomy, Rice University, Houston, Texas 77005, USA}

\author{Pengcheng Dai}

\email{pdai@rice.edu}

\selectlanguage{english}%

\affiliation{Department of Physics and Astronomy, Rice University, Houston, Texas
77005, USA}

\affiliation{Center for Advanced Quantum Studies and Department of Physics, Beijing
Normal University, Beijing 100875, China}

\begin{abstract}
Unconventional superconductivity typically emerges in the presence of quasi-degenerate ground states, and the associated intense fluctuations are likely responsible for generating the superconducting state. Here we use polarized neutron scattering to study the spin space anisotropy of spin excitations in Fe$_{1.07}$Te exhibiting bicollinear antiferromagnetic (AF) order, the parent compound of FeTe$_{1-x}$Se$_x$ superconductors. We confirm that 
the low energy spin excitations are transverse spin waves, consistent with a local-moment origin of the 
bicollinear AF order. While the ordered moments lie in the $ab$-plane in Fe$_{1.07}$Te, it takes less energy for them to fluctuate out-of-plane, similar to BaFe$_2$As$_2$ and NaFeAs.  At energies above $E\gtrsim20$ meV, we find magnetic scattering to be dominated by an isotropic continuum that persists up to at least 50 meV.  Although the isotropic spin excitations cannot be ascribed to spin waves from a long-range ordered local moment antiferromagnet, the continuum can result from the bicollinear magnetic order ground state of Fe$_{1.07}$Te being quasi-degenerate with plaquette magnetic order.  
\end{abstract}


\maketitle

Unconventional superconductivity in cuprate and heavy fermion superconductors emerge in the vicinity of multiple exotic orders that are quasi-degenerate in energy \cite{EFradkin_RMP,BKeimer_Nature,MKenzelmann_RPP,DYKim_PRX}, providing a plethora of fluctuations that may enhance or even generate superconductivity. Iron-based superconductors are found close to several different magnetic instabilities \cite{cruz,QHuang08,WBao09,WBao11,JZhao2012,JMAllred2016,MHirashi14,YSong2016_NC,SIimura,WRMeier17}, suggesting an important role for magnetism in their superconductivity \cite{dai,ScalapinoRMP}. In addition, these materials may exhibit quasi-degenerate ground states, realized through magnetic frustration and electron correlations \cite{QMSi08,QMSiNRM}. These interactions are epitomized in the iron chalcogenide FeTe$_{1-x}$Se$_x$ series, with magnetism evolving from bicollinear (BC) magnetic order in Fe$_{1+y}$Te \cite{WBao09,SLi09} towards competing stripe and N$\rm \acute{e}$el fluctuations without static magnetic order in FeSe \cite{QSWang_NC}.  Understanding the nature of magnetic fluctuations and manifestations of magnetic frustration is therefore a key step towards elucidating the physics of these materials. 

Compared to the parent compounds of iron pnictides that order at the in-plane wave vector ${\bf Q}= (0.5,0.5)$ of the paramagnetic tetragonal unit cell corresponding to the nesting wave vector of electron and hole Fermi surfaces (stripe AF order) \cite{DJSingh,IIMazin10}, the parent compound of iron chalcogenide superconductors Fe$_{1+y}$Te orders at or near ${\bf Q}=(0.5,0)$ \cite{WBao09,SLi09}, despite sharing a similar electronic structure with the iron pnictides \cite{ASubedi08,YXia09}. Furthermore, Fe$_{1+y}$Te exhibits significantly larger ordered moments \cite{WBao09,SLi09} and stronger electronic correlations \cite{ZPYin11} than iron pnictides. These results point to localized magnetism in Fe$_{1+y}$Te, although the presence itinerant carriers can cause damping of the magnetic excitations. 

At low interstitial iron concentrations ($y<0.12$), Fe$_{1+y}$Te exhibits long-range BC order with the ordering vector ${\bf Q}=(0.5,0)$ and ordered moments along the $b$-axis [Figs. 1(a) and 1(b)].
For $y\sim0.12$, a collinear short-range-ordered phase that orders at ${\bf Q}=(\delta,0)$ ($\delta\sim0.45$) with moments along $b$-axis 
is found. For $y>0.12$, helical magnetic order at ${\bf Q}=(\delta,0)$ ($\delta\sim0.38$) with moments rotating in the $bc$-plane is stabilized \cite{EERodriguez11,CStock17}. 

The complexity of magnetism in Fe$_{1+y}$Te likely arises from frustration, suggested experimentally by spin fluctuations that persist to $\sim200$ meV \cite{OLipscombe,IZaliznyak11,CStock14} compared to a much smaller Curie-Weiss temperature \cite{RHu09}. Competition between different ground states is also manifested above $T_{\rm N}$ in Fe$_{1+y}$Te exhibiting BC order, with fluctuations at an incommensurate wave vector ${\bf Q}=(\delta,0)$ shifting to the commensurate wave vector ${\bf Q}=(0.5,0)$ below $T_{\rm N}$ \cite{CStock17,DParshall12}. Theoretically, BC order is degenerate with plaquette (PQ) order that also orders at  ${\bf Q}=(0.5,0)$ \cite{HHLai17}, this degeneracy is removed through spin-lattice coupling \cite{CBBishop16} or ring-exchange \cite{HHLai17} in Fe$_{1+y}$Te, with BC order prevailing as the ground state although PQ order remains quasi-degenerate in energy. Spin fluctuations associated with the two orders are also difficult to disentangle, with measurements using unpolarized neutrons scattering interpreted as damped spin waves from BC order \cite{OLipscombe} or short-range PQ flucutations \cite{IZaliznyak11}. 
Separating spin fluctuations associated with competing states is therefore a integral part to elucidating the nature of magnetism in Fe$_{1+y}$Te. 

In this work, we study the spin space anisotropy of spin fluctuations in Fe$_{1.07}$Te exhibiting BC order below $T_{\rm N}\approx68$ K using polarized neutron scattering. We observe two transverse spin wave modes associated with the BC order that display different spin-anisotropy gaps. Although the ordered moments lie in the Fe-Te plane, spin waves corresponding to spins rotating out of the plane occur at a lower energy, similar to BaFe$_2$As$_2$ \cite{NQureshi12,CWangPRX} and NaFeAs \cite{YSong2013}. Surprisingly, we observe a continuum of isotropic scattering that extends to at least 50 meV. Our findings can be understood to result from the BC order ground state of Fe$_{1+y}$Te being quasi-degenerate with PQ order, producing an excitation spectra consisting of transverse spin waves and an isotropic spin-liquid-like response.         

\begin{figure}
\includegraphics[scale=0.65]{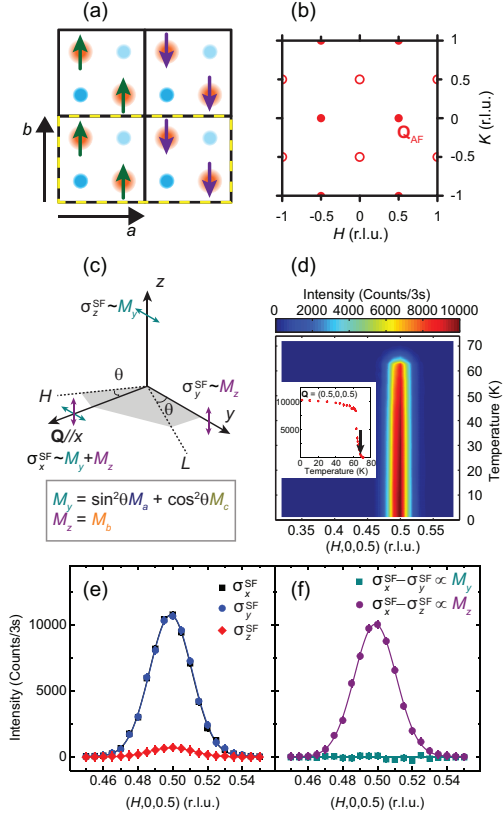} \protect\caption{(Color online) (a) The in-plane BC AF structure of Fe$_{1.07}$Te with the ordered moments along the $b$-axis. The solid black lines enclose chemical unit cells. Anti-parallel ordered moments are shown with different colors. (b) The reciprocal space of magnetically ordered Fe$_{1.07}$Te, with AF zone centers represented by red circles. For BC order domains ordering at ${\bf Q}=(0.5,0)$ (closed circle) or ${\bf Q}=(0,0.5)$ (open symbol) can form. (c) Schematic of experimental geometry, the $[H,0,L]$ scattering plane is represented by the shaded gray area, and the angle between ${\bf Q}$ and $(1,0,0)$ is $\theta$. In this scattering plane, only the domain ordering at ${\bf Q}=(0.5,0)$ is probed. (d) Color-coded temperature dependence of elastic scans along $[H,0,0.5]$ for $\sigma^{\rm SF}_{x}$ demonstrating a first-order magnetic transition with $T_{\rm N}\approx68$ K. The inset shows the temperature dependence of intensity measured at ${\bf Q}_{\rm AF}=(0.5,0,0.5)$, the arrow marks $T_{\rm N}\approx68$ K. No discernible intensity is seen at incommensurate wave vectors below $T_{\rm N}$, although above $T_{\rm N}$ there is diffuse magnetic scattering centered at an incommensurate position \cite{supplementary}. (e) Elastic scans of $\sigma^{\rm SF}_{\alpha}$ ($\alpha=x,y,z$) along $[H,0,0.5]$ at $T=2K$. (f) The differences $\sigma^{\rm SF}_x-\sigma^{\rm SF}_y$ and $\sigma^{\rm SF}_x-\sigma^{\rm SF}_z$ obtained from results in (e).}
\end{figure}

Polarized neutron scattering measurements were carried out using the IN22 triple-axis spectrometer equipped with CRYOPAD at Institut Laue-Langevin,
Grenoble, France. Heusler monochromator and analyzer with fixed $k_{\rm f}$ of 2.66 {\AA$^{-1}$ } or 3.84 {\AA$^{-1}$ } were used to carry out longitudinal polarization analysis. We aligned 7 grams of Fe$_{1+y}$Te single crystals with $y=0.07(2)$ ($a\approx b\approx3.80$ {\AA }, $c\approx6.24$ {\AA }) in the $[H,0,L]$ scattering plane, 
the amount of excess iron is estimated by comparing $T_{\rm N}\approx68$ K [inset in Fig. 1(d)] of our sample with the well-established Fe$_{1+y}$Te phase diagram \cite{EERodriguez11}. Using the tetragonal chemical unit cell of Fe$_{1+y}$Te, BC AF order is observed at ${\bf Q}_{\rm AF}=(0.5,0,L)$ with $L=0.5,1.5,2.5\dots$ [Fig. 1(b)]. Magnetic neutron scattering directly measures the magnetic
scattering function $S^{\alpha\beta}({\bf Q},E)$, which is proportional
to the imaginary part of the dynamic susceptibility $\text{Im}\chi^{\alpha\beta}({\bf Q},E)$ through the Bose
factor, $S^{\alpha\beta}({\bf Q},E)\propto[1-\text{exp}(-\frac{E}{k_{{\rm B}}T})]^{-1}\text{Im}\chi^{\alpha\beta}({\bf Q},E)$
\cite{Furrer}.  We denote the diagonal components of the magnetic scattering function $S^{\alpha\alpha}$ as $M_{\alpha}$ \cite{YSong2016}. Three neutron spin-flip (SF) 
cross sections $\sigma_{x}^{{\rm SF}}$,
$\sigma_{y}^{{\rm SF}}$, and $\sigma_{z}^{{\rm SF}}$ were measured and normalized by monitor count units (m.c.u.),
with the usual convention $x\parallel{\bf Q}$, $y\perp{\bf Q}$ in the scattering plane, and $z$ perpendicular to the scattering
plane [Fig. 1(c)]. Neutron SF cross sections measure components of $M_\alpha$ that are perpendicular to both ${\bf Q}$ and the polarization direction, therefore $M_y$ contributes to $\sigma^{\rm SF}_{x}$ and $\sigma^{\rm SF}_{z}$ whereas $M_z$ contributes to $\sigma^{\rm SF}_{x}$ and $\sigma^{\rm SF}_{y}$ [Fig. 1(c)].  Since ordered moments in Fe$_{1+y}$Te with BC order are oriented along $b$-axis which is parallel to $z$, elastic magnetic scattering should be seen in $\sigma^{\rm SF}_{x}$ and $\sigma^{\rm SF}_{y}$, as confirmed in our experiment [Fig. 1(e)]. A small peak is also observed in $\sigma^{\rm SF}_{z}$ due to non-perfect polarization of neutrons resulting in a flipping ratio of $R\approx14.5$.
$M_y$ and $M_z$ can be obtained through $M_y=c(\sigma^{\rm SF}_{x}-\sigma^{\rm SF}_{y})$ and $M_z=c(\sigma^{\rm SF}_{x}-\sigma^{\rm SF}_{z})$, with $c=(R-1)/(R+1)$. Doing so eliminates effects due to background, non-magnetic scattering and non-ideal polarization of the neutron beam \cite{CZhang2014}. For elastic magnetic scattering, a peak is seen in $M_z$ while $M_y$ is completely flat [Fig. 1(f)], as expected for BC order with moments along the $b$-axis. 

\begin{figure}
	\includegraphics[scale=0.7]{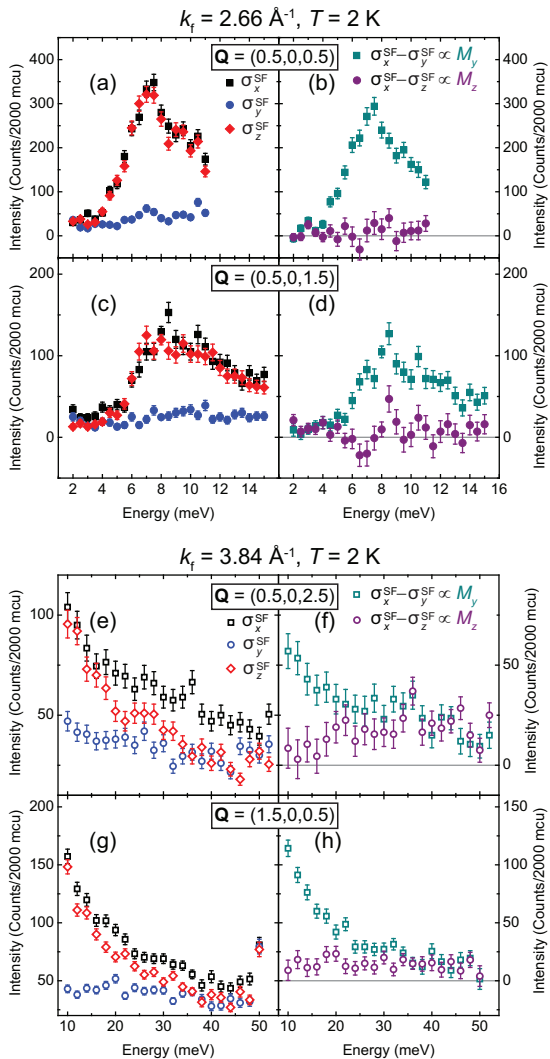} \protect\caption{(Color online) Constant-${\bf Q}$ scans of $\sigma_{x}^{{\rm SF}}$,
		$\sigma_{y}^{{\rm SF}}$ and $\sigma_{z}^{{\rm SF}}$ at (a) ${\bf Q}=(0.5,0,0.5)$, (c) ${\bf Q}=(0.5,0,1.5)$, (e) ${\bf Q}=(0.5,0,2.5)$ and (g) ${\bf Q}=(1.5,0,0.5)$. The corresponding differences $\sigma^{\rm SF}_{x}-\sigma^{\rm SF}_{y}$ and $\sigma^{\rm SF}_{x}-\sigma^{\rm SF}_{z}$ are respectively shown in (b), (d), (f) and (h). Closed and open symbols are measured with fixed $k_{\rm f}=2.66$ \AA$^{-1}$ and $k_{\rm f}=3.84$ \AA$^{-1}$, respectively.}
\end{figure}

In Fe$_{1.07}$Te, $M_z$ is uniquely associated with the direction of the ordered moments (longitudinal direction), while $M_y$ is a combination of the two transverse directions [Fig. 1(c)]. This contrasts with similar setups in BaFe$_2$As$_2$ \cite{CWangPRX} and NaFeAs \cite{YSong2013} where $M_z$ corresponds to a transverse direction and $M_y$ is a mixture of the longitudinal direction and another transverse direction. Therefore for Fe$_{1+y}$Te, fluctuations along the longitudinal direction can be directly probed in $M_z$. 

Fig. 2 summarizes constant-${\bf Q}$ scans at several equivalent wave vectors corresponding to AF zone centers. Whereas at low energies the magnetic fluctuations are dominated by transverse spin waves in $M_y$ [Fig. 2(a)-(d)] \cite{CStock17}, clear longitudinal fluctuations
are seen in $M_z$ above $\sim20$ meV and the excitations become isotropic with $M_y\approx M_z$ above $\sim35$ meV [Fig. 2(e)-(h)]. Isotropic scattering that appears for $E\gtrsim$ 20 meV, as indicated by the broad onset of longitudinal fluctuations, depends weakly on energy and extends over a large energy range (persisting up to at least 50 meV), forming a continuum of scattering. Measurement of non-spin-flip cross sections confirm these conclusions \cite{supplementary}. Such isotropic excitations is unexpected for an ordered local moment antiferromagnet which should exhibit transverse spin waves, and also cannot be accounted for by Fermi surfaces that are connected by ${\bf Q}=(0.5,0.5)$ \cite{ASubedi08,YXia09}. Instead, as discussed below, the isotropic continuum of scattering can be identified as fluctuations associated with PQ order that is quasi-degenerate with the BC ground state \cite{HHLai17}. 

Since the two transverse directions are mixed in $M_y$ depending on the angle between ${\bf Q}$ and $H$ [Fig. 1(c)], measurements at equivalent wave vectors are needed to separate them \cite{CZhang2014}. Combining data from equivalent wave vectors from Fig. 2, $M_a$, $M_b$ and $M_c$ can be obtained, as shown in Fig. 3(a). For $M_a$ and $M_c$ corresponding to the two transverse directions, spin wave modes exhibiting different anisotropy gaps can be clearly seen, along with a continuum of isotropic scattering at higher energies. Although the ordered moments are along 
the $b$-axis inside the $ab$-plane, the $c$-axis polarized spin waves are lower in energy similar to iron pnictide parent compounds \cite{CWangPRX,YSong2013}. The low-energy transverse spin waves also display a dispersion of $\sim5$ meV along $L$ \cite{supplementary}, in agreement with previous results \cite{CStock14}.

The $c$-axis polarized spin waves dominating for $E\lesssim10$ meV can also be seen in $L$-scan of $M_y$ in Fig. 3(b), the fast drop of intensity with increasing $L$ is due to the decreasing contribution of $M_c$ as ${\bf Q}$ turns towards the $c$-axis [Fig. 1(c)]. This should be contrasted with isotropic paramagnetic scattering above $T_{\rm N}$ with $\sigma^{\rm SF}_y\approx\sigma^{\rm SF}_z$ [Fig. 3(c)], which falls off with $L$ following the magnetic form factor [Fig. 3(b)]. The $L$-dependence of $M_y$ at 2 K in Fig. 3(b) is fit to a lattice sum of Lorentzians that has both the $c$- and $a$-axis polarized components, resulting in a ratio of 4(1) for the two components, consistent with Fig. 3(a). The strongly anisotropic magnetic excitations shown in Fig. 3(a) suggests spin anisotropy may affect calculation of the local susceptibility at low energies and application of the sum rule, where isotropic scattering is typically assumed \cite{dai,GXu}. Previously, it has been suggested that the strong peak in energy at ${\bf Q}_{\rm AF}=(0.5,0)$ and $E\sim7$ meV in Fe$_{1+y}$Te may be linked to the resonance seen in superconducting Fe$_{1+y}$Te$_{1-x}$Se$_x$ that occurs at a similar energy but different wave vector ${\bf Q}=(0.5,0.5)$ \cite{CStock11,IZaliznyak11}. Here we establish that the strong peak in Fe$_{1.07}$Te is polarized along the $c$-axis similar to the resonance mode in FeSe \cite{MMa17_PRX}, but different from the resonance mode in superconducting Fe$_{1+y}$Te$_{1-x}$Se$_x$ that has both in-plane and out-of-plane components \cite{PBabkevich11,KProkes12}. Our observation that the $c$-axis polarized spin waves being lower in energy for Fe$_{1+y}$Te with BC order also accounts for rotation plane ($bc$-plane rather than $ab$-plane) of the helical magnetic structure seen in samples with $y>0.12$ \cite{EERodriguez11,CStock17}. 

\begin{figure}
	\includegraphics[scale=0.75]{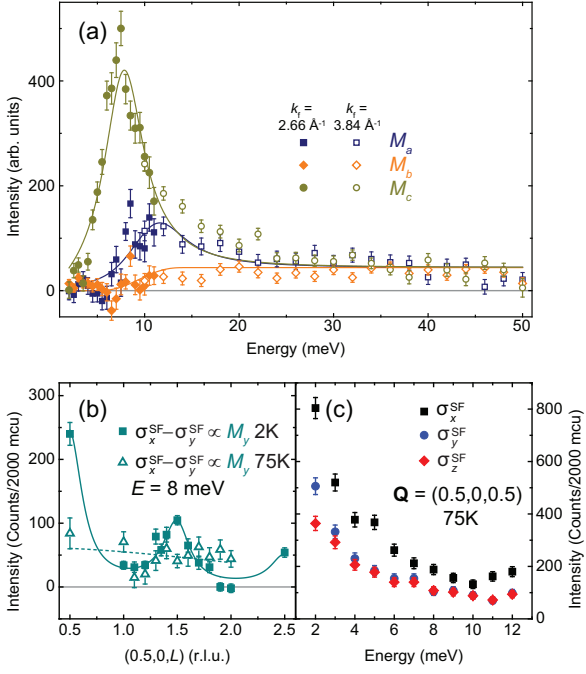} \protect\caption{ (Color online) (a) $M_a$, $M_b$ and $M_c$ for the AF zone centers obtained from data in Fig. 2. Results obtained using different $k_{\rm f}$ are scaled for best match at $E=10$ meV. The solid lines are fits to damped harmonic oscillators in $M_a$ and $M_c$, and a broad isotropic response appears in all three channels. (b) Constant-energy scans of $M_y$ along $(0.5,0,L)$ for $E=8$ meV at 2 K and 75 K. The solid line is fit to a lattice sum of Lorentzian peaks, the dashed line represents $L$-independent isotropic scattering that is only modulated by the Fe$^{2+}$ magnetic form factor. (c) Constant-${\bf Q}$ scans of the three SF cross sections in the paramagnetic state ($T =75$ K) at ${\bf Q}=(0.5,0,0,5)$. Anisotropy is only observed for $E\lesssim2$ meV, extending down to $E=0$ forming anisotropic diffuse magnetic scattering \cite{supplementary}. 
		}
\end{figure}

\begin{figure}
\includegraphics[scale=0.75]{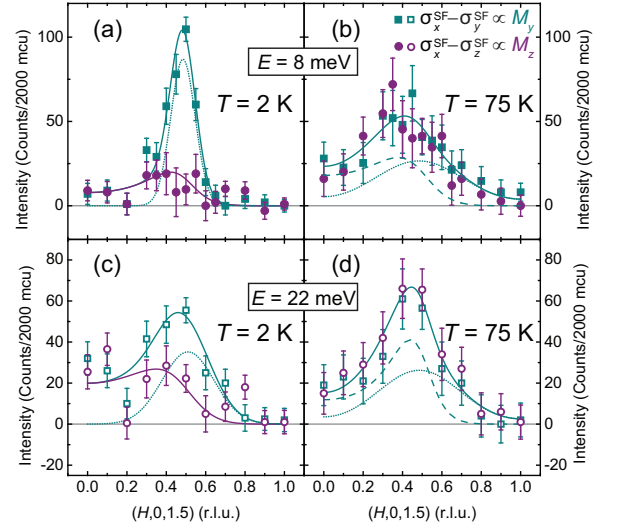} \protect\caption{(Color online) Scans of $\sigma^{\rm SF}_x-\sigma^{\rm SF}_y$ and $\sigma^{\rm SF}_x-\sigma^{\rm SF}_z$ along $[H,0,1.5]$ for $E=8$ meV at (a) 2 K and (b) 75 K. Similar scans are shown for $E=22$ meV at (c) 2 K and (d) 75 K. 
The closed and open symbols are measured with fixed $k_{\rm f}=2.66$ \AA$^{-1}$ and
$k_{\rm f}=3.84$ \AA$^{-1}$, respectively. Lines are fits as described in the text.
}
\end{figure}

Having established the presence of both transverse spin waves and an isotropic continuum of scattering at ${\bf Q}_{\rm AF}$, we studied the momentum dependence of these excitations in comparison with isotropic paramagnetic scattering above $T_{\rm N}$, as shown in Fig. 4 (temperature evolution of the scattering cross sections is shown in Supplemental Materials \cite{supplementary}). For $T=2$ K [Figs. 4(a) and (c)], the momentum dependence of $M_z$ can be described as short-range PQ correlations \cite{IZaliznyak11,IZaliznyak15}, and $M_y$ as a sum of the same short-range PQ correlations and a Gaussian peak centered at ${\bf Q}=(0.5,0)$ (dotted lines). These results provide additional evidence that below $T_{\rm N}$, the isotropic scattering that appears in both $M_y$ and $M_z$ is associated with PQ order, whereas the signal only present in $M_y$ is due to transverse spin waves of the BC ground state. Below $T_{\rm N}$, the transverse spin waves dominate for $E=8$ meV [Fig. 4(a)] whereas for $E=22$ meV the two components become comparable [Fig. 4(c)].  Above $T_{\rm N}$, the scattering becomes isotropic and centered at an incommensurate position ($\sim$0.4,0) [Figs. 4(b) and (d)], and can also be described as a sum of short-range PQ correlations (dashed lines) and a Gaussian peak centered at ${\bf Q}=(0.5,0)$ (dotted lines). Our results suggest above $T_{\rm N}$ fluctuations associated with BC and PQ orders are both present, with the overall intensity centered at ${\bf Q}\sim(0.4,0)$. When BC order is selected as the ground state below $T_{\rm N}$, transverse spin waves become dominant at low energies and the overall signal shifts to ${\bf Q}=(0.5,0)$, as experimentally observed \cite{CStock17,DParshall12}. 

The isotropic continuum of scattering in Fe$_{1.07}$Te is clearly inconsistent with transverse spin waves arising from BC order, it also cannot be interpreted as two-magnon scattering which only appears along the longitudinal direction \cite{WSchweika02,THuberman05}. 
Instead, it is most naturally associated with short-range PQ order \cite{HHLai17.2}: 
while a long-range PQ order would generate spin waves which appear
only in the transverse channels, a short-range PQ order produces collective excitations that are isotropic.
The quasi-degeneracy \cite{HHLai17} of the short-range PQ order with the long-range BC order 
ensures that such excitations occur at relatively low energies, as we have observed here. The presence of both spin waves and a continuum of scattering is also observed in proximate spin liquid materials such as KCuF$_3$ \cite{BLake2005}
and $\alpha$-RuCl$_3$ \cite{ABanerjee}, where weak magnetic order is the ground state.
Spin excitations in these materials are from quasi-degeneracy of spin liquid states and magnetically ordered states, with
spin waves from ordered state appear at lower energies \cite{ABanerjee}. In Fe$_{1+y}$Te below $T_{\rm N}$, the similar observation is caused by quasi-degeneracy of two different magnetic orders. 

The picture of PQ order being quasi-degenerate with BC order also implies that a small external perturbation can tilt the balance in the stability of the two orders. Indeed, it was found that the large magnetic moment on interstitial iron in Fe$_{1+y}$Te$_{0.62}$Se$_{0.38}$ induces short-range spin arrangements resembling the PQ order \cite{VThampy12}, suggesting excess interstitial iron in Fe$_{1+y}$Te would similarly favor PQ over BC order locally. This view is consistent with the observation that BC order is destabilized with increasing excess iron \cite{EERodriguez11}. 

To summarize, our polarized neutron scattering results in Fe$_{1.07}$Te point to the presence of both transverse spin waves associated with BC  order and a continuum of isotropic excitations likely associated with short-range PQ order. This provides evidence for the quasi-degeneracy between the short-range PQ order and the long-range BC order and, thereby, the strongly frustrated nature of local-moment magnetism in the iron chalcogenides. Our findings underscore the importance of electron correlations to the magnetism and superconductivity in the iron-based materials.

The neutron scattering work at Rice is supported by the U.S.
DOE, BES under contract no. DE-SC0012311 (P.D.). A part of the materials work at Rice
is supported by the Robert A. Welch Foundation Grants No. C-1839 (P.D.).
The theory work at Rice is supported by the NSF Grant No.\ DMR-1611392 and the Robert A.\ Welch Foundation Grant No.\ C-1411 
(W.-J.H., H.-H.L. and Q.S.) and the NSF Grant No.\ DMR-1350237 (H.-H.L. and W.-J.H.),  and
a Smalley Postdoctoral Fellowship of the Rice Center for Quantum Materials (H.-H. L.).

\end{document}